# Optically Isotropic and Monoclinic Ferroelectric Phases in PZT Single Crystals near Morphotropic Phase Boundary

Alexei A. Bokov, Xifa Long, and Zuo-Guang Ye

*Department of Chemistry and 4D LABS, Simon Fraser University, Burnaby,*

*British Columbia, V5A 1S6, Canada*

We report the finding of unusual scale-dependent symmetry below the ferroelectric Curie temperature in the perovskite Pb(Zr$_{1-x}$Ti$_x$)O$_3$ single crystals of morphotropic phase boundary compositions. The crystals of tetragonal symmetry (from x-ray diffraction experiments) on sub-micrometer scale exhibit a macroscopic (optically determined) cubic symmetry. This peculiar optical isotropy is explained by the anomalously small size of tetragonal ferroelectric domains. Upon further cooling the crystals transform to the phase consisting of micrometer-sized domains of monoclinic *Cm* symmetry.

Ferroelectric (FE) properties attractive for applications are mostly observed in solid solutions with the composition close to the morphotropic phase boundary (MPB), which corresponds to the solution concentration separating different FE phases. Typical examples include the ceramics of Pb(Zr$_{1-x}$Ti$_x$)O$_3$ solid solutions (PZT), which are currently the most widely used piezoelectric materials, and the new generation piezoelectric crystals based on perovskite relaxor-ferroelectrics, such as (1-$x$)Pb(Mg$_{1/3}$Nb$_{2/3}$)O$_3$-$x$PbTiO$_3$ (PMNT) [1]. While remarkable technological progress has been achieved in the last decades, the relation between the crystal structure and the properties in the MPB region is far from being understood. In early literature [1,2] the MPB in PZT and relaxor-based materials was reported to separate the rhombohedral (space group *R3m*) and tetragonal (*P4mm*) phases. Investigations of MPB have been significantly advanced by Noheda *et al.* [3] who discovered an intermediate monoclinic (*Cm*) phase in the PZT ceramics using high-resolution synchrotron x-ray diffraction. At room temperature, the rhombohedral (R), monoclinic (M) and tetragonal (T) phases were found to exist in the composition ranges of $x < 0.45$, $0.45 < x < 0.48$ and $x > 0.48$, respectively. Later, similar intermediate M phase was found in the relaxor-based solid solutions [4] and the role of this phase in the enhancement of piezoelectric properties was widely discussed.[1]

A remarkable peculiarity of the PZT and relaxor-based solid solutions is that the crystal symmetry may depend on the length scale under consideration. In particular, Glazer *et al.* [5] suggested that at the local (several unit cells) level the symmetry of PZT is M for all compositions, but on the larger length scale of x-ray or neutron diffraction probe (~100 nm), it is M only within the composition range the MPB. Alternatively, it was shown theoretically [6] that the structure of lamella nanodomains (which was observed near the MPB in PZT [7] and PMNT [8] by means of transmission electron microscopy) can give rise to the x-ray diffraction pattern characteristic of M phase, even though the symmetry of the individual domains is R or T. The very existence of M phase was thereby questioned.

The macroscopic symmetry, which determines the crystal physical properties, can only be established optically in good-quality single crystals. Growth of PZT crystals with MPB compositions has been a challenge and they are not widely available. In the early polarizing microscopy work by Fesenko *et al.* [9] only the R and T phases were reported in PZT crystals around MPB composition and the M phase was not found, probably because it was not expected at that time. On the other hand, it cannot be easily distinguished optically from the mixture of R and T phases.

In the present work we have observed the M phase in the PZT crystals. Upon heating, this phase transforms into another FE phase which exhibits highly unusual behavior, namely the phase is found to be optically isotropic.

PZT single crystals were grown by a top-seeded solution growth technique which was developed recently in our laboratory. The perovskite-type structure was confirmed by standard X-ray diffraction. The composition of specimens was verified using energy dispersive X-ray spectroscopy (EDAX analyzer of Strata DB235 scanning electron microscope). Dielectric and optical investigations were made using Solartron 1260/1296 dielectric spectrometer and Olympus BX60 polarizing microscope equipped with Berek and Sénarmont compensators. A Linkam THMS600 optical heating stage was used for high-temperature observations. For dielectric measurements, gold electrodes were sputtered.

The temperature dependence of the dielectric constant of the PZT crystal with $x \approx 0.45$ is shown in Fig. 1. Two dielectric anomalies, namely, a diffuse step at about 100 $^{o}$C and a sharp maximum at 390 $^{o}$C, are observed. According to phase diagram obtained by x-ray diffraction in ceramic samples [3] these anomalies can be attributed to the transition from ferroelectric M to T phase separated by the inclined MPB and to the Curie point ($T_C$), respectively..

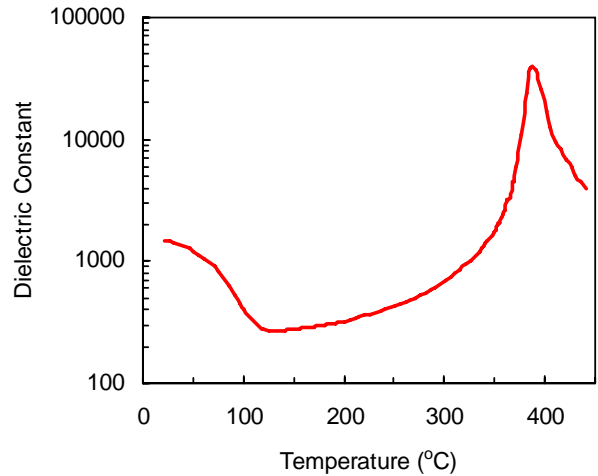

FIG. 1. Temperature dependence of dielectric constant at 100 kHz in Pb(Zr$_{1-x}$Ti$_x$)O$_3$ crystal with $x \approx 0.45$

To discriminate between different phases we analyze the extinction positions and the directions of domain walls in the crystal plates having large faces represented by the (001) planes (we refer to the crystallographic axes of the paraelectric cubic phase). Generally the deformation of the paraelectric perovskite structure ($Pm\bar{3}m$) below $T_C$ can result [see Fig. 2(a)] in T, R, orthorhombic *Amm*2 (O), monoclinic *Pm* ($M_C$), two types of monoclinic *Cm* ($M_A$ and $M_B$) or triclinic (Tr) symmetry (all these phases were observed or suspected previously in different FE perovskites around MPB [1,10]). The domain viewed under crossed polars appears to be in extinction when the (mutually perpendicular) vibration directions of the polarizer and analyzer are directed along the (mutually perpendicular) slow and fast vibration directions of the domain (i.e. major and minor axes of the ellipse formed by cross-sectioning of the optical indicatrix by the plane of the crystal plate). In optically uniaxial R and T phases the symmetry axis of optical indicatrix collinear to spontaneous polarization ($P_S$) is aligned along one of the <111> (in R phase) or <100> (in T phase) crystallographic directions. Accordingly, the vibration directions of all the domains in the (001) plate (which is studied) are almost parallel or perpendicular to one another and the extinction angle ($\theta$) determined as the smaller angle between the domain's vibration direction and the [100] crystallographic axis (with the larger angle being $90° + \theta$) is almost the same for all domains, namely $\theta = 45°$ in the R phase and $\theta = 0°$ in the T phase [see Fig. 2(b)]. Therefore, all domains must show extinction simultaneously. In the optically biaxial O phase one of the indicatrix axes is parallel to the <100> direction and two others are parallel to <110>, so that different domains in the (001) platelet may appear in extinction at $\theta = 45°$ or at $\theta = 0°$. The restrictions imposed by symmetry in the optically biaxial M phases are less severe: one of the indicatrix axes should be perpendicular to the mirror plane of the unit cell, i.e. (110) plane in the *Cm* phases ($M_A$ and $M_B$) or (100) plane in the $M_C$ phase, and the other two axes can be anywhere in the mirror plane. As a result, some domains within the (001) platelet may extinguish at $\theta = 45°$ in the *Cm* phases or $\theta = 0°$ in the $M_C$ phase, and the extinction angles of other domains may have two different values: $\theta$ or $-\theta$, where $\theta$ is *a priori* unknown and may depend on composition and temperature. At the same time, the values of $\theta = 0°$ and $\theta = 45°$ are not expected in the *Cm* and $M_C$ phases, respectively. Finally, in the Tr phase, where the position of the indicatrix is not restricted by symmetry, six different extinction angles ($\theta_1$, $\theta_2$, $\theta_3$, $-\theta_1$, $-\theta_2$, and $-\theta_3$) are possible in the (001) platelet, but neither $\theta = 0°$ nor $\theta = 45°$ are expected. In the O, M and Tr crystals or in the MPB crystals containing domains of different symmetry (e. g. T and R), the situation is likely [while impossible in the case of pure R or T (001) platelets] where no extinction is observed at all. This should happen in those areas of the platelet where two or more domains with nonparallel vibration directions overlap on the way of propagating light.

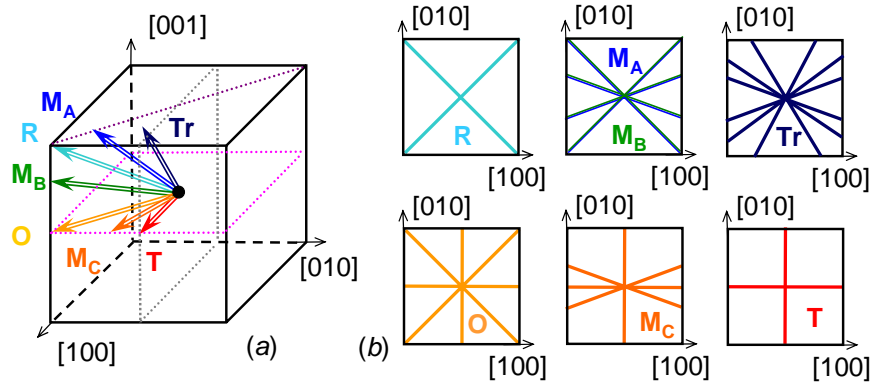

FIG. 2. Schematics for different FE phases: (a) $P_S$ directions with respect to cubic unit cell of paraelectric phase and (b) possible extinction positions in (001) crystal plate

The photographs of the PZT crystal with $x \approx 0.45$ are shown in Fig. 3. Irregular patterns of interference colors suggest the presence of complex structure of domains with the size much smaller than the crystal thickness. When the crossed polars are in <100> or <110> positions [Figs. 3(a) and 3(b), respectively] the crystal remains bright, which, as discussed above, precludes the existence of large homophase regions of R or T symmetry. On the other hand [Fig. 3(c)], some comparatively large areas appear to be in extinction (black) when $\theta \approx 36°$ (this angle may vary by several degrees in different parts of the crystal, probably due to a slight variation of composition $x$). This extinction angle is permissible in the M or Tr phases, but not in the T, R or O phases. Small extinction areas are also occasionally observed with $\theta=45°$ [Fig. 3(b)], which is not expected for the $M_C$ or Tr structure. Therefore, the observed extinction positions are fully compatible with the monoclinic $Cm$ phase, in agreement with the x-ray diffraction data [3]. The large regions without extinctions can be due to superimposed $Cm$ domains having different vibration directions. Note, however, that we cannot preclude the possibility of small inclusions of a different phase in these regions.

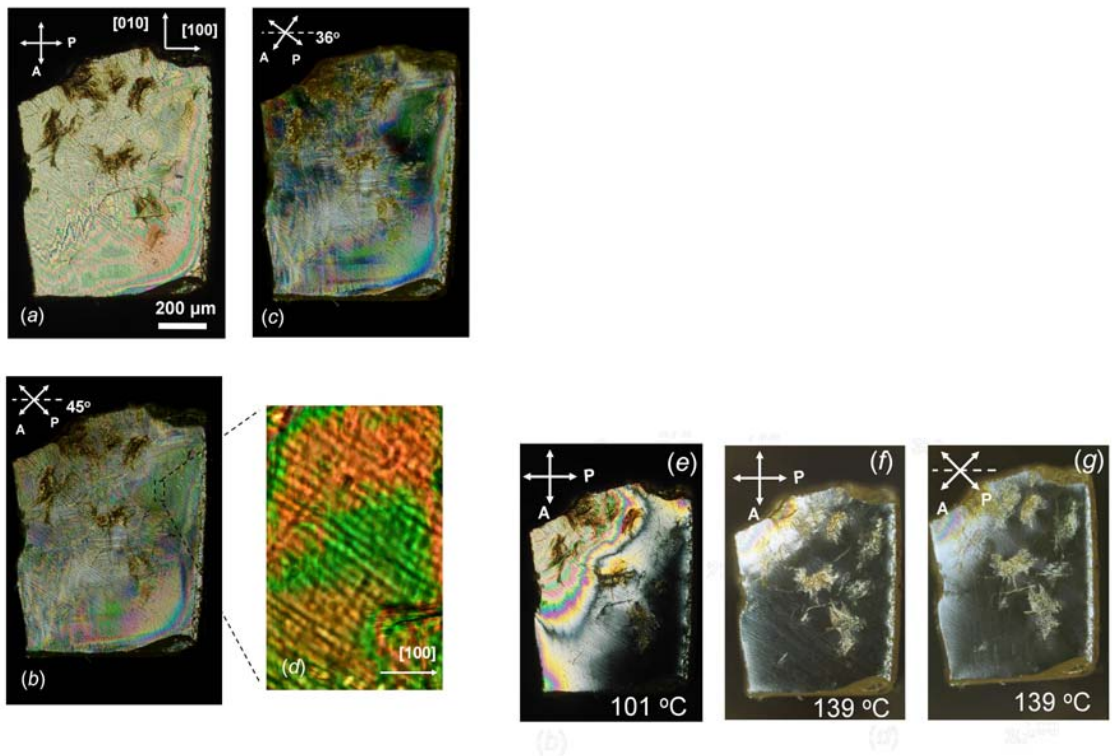

FIG. 3 Micrographs of (001) plate (0.12 mm thick) of Pb(Zr$_{1-x}$Ti$_x$)O$_3$ crystal with $x \approx 0.45$ at different positions of crossed polars (shown by arrows) and temperatures: (a-d) room temperature, (e) 101 °C and (f,g) 139 °C. Enlarged view in (d) shows the FE domains. Pseudocubic crystallographic directions are indicated.

The other solid confirmation for the M symmetry is the domain walls direction. In many regions the domains are large enough to be examined under microscope, as illustrated in Fig. 3(d) where lamella 2-3 μm thick domains are clearly visible. The domain walls are almost perfectly straight, which suggests that they are non-180° walls or 180° walls with $P_S$ parallel to the crystal surface. The walls form an angle of 32 - 37° (different in different parts of the crystal) with one of the <100> directions. These angles are impossible in the T, R or O phases in which the wall's traces on the (001) plane (crystal surface) must be parallel to the <100>

or <110> directions due to the elastic and electric compatibility requirements [11]. On the contrary, in the M phases other directions of walls (depending on composition and temperature) are possible [11] and really observed [12].

Unusual behavior is found in the T phase. This phase is characterized by a significant tetragonality ($c/a$) value of about 1.03 [3]. In other perovskite-type ferroelectrics a comparable spontaneous deformation gives rise to a large birefringence of $10^{-1} - 10^{-2}$ [13]. On the contrary, our crystals are found to be almost isotropic in the T phase. Fig. 3(e) shows the crystal at the temperature of the M-T phase transition (101 $^{o}$C) where the top-left part of the specimen is still in the M phase while the bottom-right part is already in the T phase. The phase front moves gradually upon heating so that at 139 $^{o}$C [Fig. 3(b,c)] only a small portion of the crystal remains in the M phase. As confirmed by the EDS analysis, this spreading of the phase transition temperature is due to the spatial variation of $x$. In the T phase the crystal remains dark at any positions of crossed polars [compare Figs. 3(f) and 3(g)]. More specifically, very small optical retardation is measured (by compensator) in some regions of the crystal, which corresponds to a birefringence of less than $10^{-3}$; in most other regions the birefringence is smaller than the detection limit of the compensator ($\sim 10^{-6}$). The birefringence in the birefringent regions vanishes above $T_C$. The extinction in these regions is observed in the <100> directions (i.e. $\theta = 0^o$), in agreement with the T symmetry. The particular (e.g. slow) vibration direction is parallel to the [100] direction in some areas and perpendicular to the same direction in some others. No distinct domain boundaries are observed: the spatial variation of the birefringence is gradual. Similar behavior (darkness at any position of crossed polars) is observed in other viewing directions [non-(001) crystal plates]. We do not notice correlation between the crystal thickness and measured retardation. However, a much larger (practically "normal") birefringence is observed in the T phase of some crystals.

The observed optical isotropy of the microscopically tetragonal (according to x-ray diffraction) phase can be related to the polydomain state with a very small size of FE domains. The total retardation for the light passing through many superimposed T domains in a (001) platelet can be calculated as $\Delta n \left( \sum_i d_i - \sum_j d_j \right)$, where $d_i$ and $d_j$ are the thicknesses of the $a$-domains in which slow vibration direction is parallel and perpendicular to the [100] direction of the whole crystal, respectively, and $\Delta n$ is the "intrinsic" birefringence in a single domain (which is comparatively large). When formed upon cooling through $T_C$ the domains in a defect-free crystal have equal probabilities to develop with $P_S$ along one of the <100> directions. If the $P_S$ directions in different domains are random and their size is much smaller than the crystal size, the relation $\sum_i d_i = \sum_j d_j$ is satisfied, leading to zero optical retardation, i.e. on the average, the crystal behaves like an isotropic medium.

When the incident ray is not parallel to [001], the vibration directions of the superimposed domains may not be directed at right angle to each other. However, additional retardation produced by every successive domain on the path of the ray is also randomly positive or negative. Therefore, the domains compensate one another and the total retardation is also expected to be zero.

However, the domain structure is influenced by many factors including domain walls energy, depolarization fields, crystal surface etc. and in normal FE crystals the domains are usually comparatively large and/or preferably oriented so that the light ray passes through few domains having different directions of optical indicatrix.[14] As a result, the crystals are typically highly anisotropic even in a polydomain state. Nevertheless, due to domain compensation effect, the measured birefringence can still be significantly reduced as compared to its "intrinsic" value [14].

In PZT ceramics of near-MPB composition, nanoscale domains were observed by transmission electron microscopy [7,10]. Therefore, the compensation and averaging effects described above are possible in the crystals of ~100 μm thickness which we study. The scattered areas of large birefringence can be the regions of occasionally large domains in which the averaging is not perfect, or can be attributed to the influence of crystal imperfections and related internal stresses leading to the preferred orientation of the domains. The fact that extinction positions in these regions are parallel to the <100> directions confirms that the symmetry of sub-micrometer domains is T and their $P_S$ vectors are along <100>.

The effect of optical isotropy similar to the one we found in the T phase of PZT crystals was previously observed in FE phase of relaxors where domains are known to be tiny, e.g. in PMNT solid solutions with comparatively small PT content [15] or in $Pb(Zn_{1/3}Mb_{2/3})O_3$ crystals [15,16] whose the domains size is in the range of 40 - 200 nm [17]. The nanosize of FE domains in relaxors and their random directions are related to the fact that they form from polar nanoregions existing in ergodic relaxor phase (intermediate between paraelectric and FE phases) [18]. Polar nanoregions also determine other properties specific of relaxors, such as diffuseness of the peak in the temperature dependence of dielectric constant around $T_C$. In PZT crystals, the paraelectric phase transforms directly into an FE one via a sharp transition (see Fig. 1) and polar nanoregions typical of relaxors are not expected to exist. So the origin of the anomalously small FE domains and their directional randomness in PZT single crystals remains to be elucidated.

According to Rossetti *et al.* [19], thermodynamics predicts, for the MPB compositions, spherical degeneration of the polarization directions and drastic decrease of the domain wall energy, which could explain the observed miniaturization of the domain structure in PZT. They also suggested that the decoupling of the polarization from the crystal lattice leads to the formation of a polar glass state with the properties similar to those of amorphous ferromagnets. However, the latter suggestion can hardly be reconciled with the fact that the vibration directions of the birefringent regions in T phase are exactly along <100>. Alternatively the behavior can be explained in terms of the theoretical approach by Imry and Ma [20] who showed that in the systems with a second-order phase transition and rotational degeneration of the order parameter, the ordered (e.g. FE) state is unstable due to the influence of *arbitrary small* random fields, so that the system breaks up into small domains of size $L$ and the crystal becomes isotropic on the scale much larger than $L$. In PZT the random local electric fields can be associated with the quenched disorder of Ti and Zr cations. The weakly-birefringent regions appear as a result of partial stabilization of FE state by the large-scale elastic inhomogeneities.

In conclusion, we found that in PZT crystals of MPB composition the symmetry of the phase existing below $T_C$ depends on the scale under consideration: on the sub-micrometer scale it is known to be tetragonal (from x-ray diffraction experiments), while the macroscopic (optically determined) symmetry is cubic. Such a kind of symmetry discrepancy is known for relaxors but is unusual for normal ferroelectrics such as PZT. It is explained by the anomalously small size of the tetragonal domains with the $P_S$ vectors randomly distributed over the <100> directions. This peculiar domain structure is supposed to be caused by quenched random fields as predicted theoretically [20]. At lower temperatures the crystals are found to transform to another phase with macroscopic FE domains whose behavior implies the monoclinic *Cm* symmetry, consistent with the x-ray diffraction results.

The work is supported by the Office of Naval Research (Grant No. N00014-06-1-0166) and the Natural Science & Engineering Research Council of Canada.